\documentclass[prd,showpacs,preprint]{revtex4}
\usepackage{amsmath}
\usepackage{graphicx}
\usepackage{braket}

\begin{document}

\title{Contributions from SUSY-FCNC couplings to the interpretation of the HyperCP events for the decay
$\Sigma^+ \rightarrow p \mu^+ \mu^-$}
\author{Gao Xiangdong}
\email{gaoxiangdong@pku.edu.cn}
\author{Chong Sheng Li}
\email{csli@pku.edu.cn}
\author{Zhao Li}
\email{zhli.phy@pku.edu.cn}
\author{Hao Zhang}
\email{haozhang.pku@pku.edu.cn} \affiliation{Department of Physics,
Peking University, Beijing 100871, China}

\date{\today}

\begin{abstract}
The observation of three events for the decay $\Sigma^+ \to p \mu^+ \mu^-$
with a dimuon invariant mass of $214.3\pm0.5$MeV by the HyperCP collaboration
imply that a new particle X may be needed to explain the observed dimuon
invariant mass distribution. We show that there
are regions in the SUSY-FCNC parameter space where the $A^0_1$ in the NMSSM
can be used to explain the HyperCP events without contradicting all
the existing constraints from the measurements of the kaon decays, and
the constraints from the $K^0-\bar{K}^0$ mixing are
automatically satisfied once the constraints from kaon decays are
satisfied.
\end{abstract}


\maketitle

\section{introduction}
Recently, there are a great deal of interest
\cite{Gorbunov:2005nu,Geng:2005ra,Deshpande:2005mb,He:2005we,Dukes:2006db,Demidov:2006pt,He:2006fr,He:2006uu,Chen:2007uv,Mangano:2007gi}
in the interpretation of the observed three events for the decay
$\Sigma^+ \rightarrow p \mu^+ \mu^-$ with a dimuon invariant mass of
$214.3\pm0.5$MeV, which were reported by the HyperCP Collaboration
\cite{Park:2005ek}. The branching ratio based on the three events
for this process is $(8.6^{+6.6}_{-5.4}\pm5.5) \times 10^{-8}$
\cite{Park:2005ek}. It has been argued that in the framework of the
Standard Model(SM) it is possible to account for the total branching
ratio when the long-distance contributions are properly included,
but all the three events are around 214MeV cannot be explained
\cite{Park:2005ek,He:2005yn}. If no new evidence support the SM
explanations in the future experiments with more events, it is most
likely to interpret the three events with the existence of a new
particle, X, beyond the SM. However, a new particle explanation for
the HyperCP events seems too radical because there are not earlier
experiments which observe such a 214MeV new particle. If this new
light particle does indeed exists and contributes to the hyperon
decay, it may also contribute to the kaon and B-meson decays. So,
the fact that lots of experiments at this low-energy region did not
observe the new 214MeV particle means that strong constraints has
been imposed on the new particle explanation for the HyperCP events.

The authors of Ref. \cite{He:2006uu} proposed a argument to explain
the HyperCP events for the hyperon decay with the new particle X
without contradicting with the constraints on the X from the
low-energy experiments. As shown in Ref. \cite{He:2006uu}, in
addition to the flavor-changing two-quark contributions, there are
also four-quark contributions arising from the combined effects of
the usual SM $|\Delta S| = 1$ operators and the flavor-conserving
couplings of X, which are comparable with the two-quark ones and
cancel sufficiently to lead to suppressed rare kaon decays rates
while combining the above two kinds of contributions yields $\Sigma
\to p \mu^+ \mu^-$ rates within the required bounds.

Based on the analysis in Ref. \cite{He:2006uu}, the authors of Ref.
\cite{He:2006fr} pointed out that a light pseudoscalar Higgs
particle $A^0_1$ in the next-to-minimal supersymmetric standard
model (NMSSM) \cite{Drees:1988fc,Ellis:1988er,Franke:1995tc} can be
identified with X. In fact, the mass of the light pseudoscalar Higgs
particle $A^0_1$ in the NMSSM can be as small as 214MeV in the
large-$\tan \beta$ limit. Under some assumptions, it has been shown
in Ref.~\cite{He:2006fr} that there are regions in the parameter
space where $A^0_1$ can satisfy the following constraints:
\begin{eqnarray}
\label{formula:constraint}
\mathcal{B}(K^{\pm} \rightarrow \pi^{\pm}
A^0_1) & \lesssim & 8.7 \times 10^{-9}, \nonumber \\
\mathcal{B}(K_s \rightarrow \pi^0
A^0_1) & \lesssim & 1.8 \times 10^{-9}, \nonumber \\
\mathcal{B}(B \rightarrow X_s A^0_1) & \lesssim & 8.0 \times
10^{-7},
\end{eqnarray}
which are obtained in Ref. \cite{He:2006uu} from the measurements of
the kaon and B-meson decays
\cite{Ma:1999uj,Park:2001cv,Batley:2004wg,Aubert:2004it,Iwasaki:2005sy},
and simultaneously explain the HyperCP events.

However, the author of Ref. \cite{He:2006fr} only considered
contributions from the SUSY charged current, i.e.,contributions
arising from the exchanges of chargino and squark, and do not
include contributions arising from SUSY-flavor-changing neutral
currents(FCNC). It is well known that the SUSY-FCNC couplings can
yield important(and, sometimes, even dominate) contributions to
low-energy flavor physics, so further investigation on the
possibility of the SUSY-FCNC mediating HyperCP events is needed. In
this paper, we show that the SUSY-FCNC effects also can explain the
HyperCP events and satisfy all the constraints in
Eq.(\ref{formula:constraint}). We adopt the mass insertion
method\cite{Hall:1985dx,Gabbiani:1988rb,Hagelin:1992tc,Gabbiani:1996hi}
to parameterize the flavor-changing effects and calculate SUSY-FCNC
contributions to branching ratio of $\Sigma \rightarrow p A^0_1$ and
rare kaon decays. This method introduce the super-CKM basis for the
quark and squark states. The couplings of quarks and squarks to the
neutral gauginos are flavor diagonal, while the flavor-changing SUSY
effects are exhibited in the off-diagonal terms of the squark mass
matrix denoted by $(\Delta^{q}_{ij})_{IJ}$, where $I, J = L, R$ and
$i, j = 1, 2, 3$ indicate chiral and flavor indices respectively,
and $q = u, d$ denote the type of quark. The squark propagator is
then expanded as a series of $(\delta^{q}_{ij})_{IJ} =
(\Delta^{q}_{ij})_{IJ} / \tilde{m}^2$, where $\tilde{m}$ is an
average squark mass. Using the mass insertion method, we can perform
calculations of the SUSY-FCNC contributions to $\Sigma \rightarrow p
A^0_1$ and rare kaon decays. Since the relevant
$(\delta^{q}_{ij})_{IJ}$ does not involve in the B-mesons decay, we
do not consider the constraints from B-mesons decay. It is well
known\cite{Hagelin:1992tc,Gabbiani:1996hi,Hagelin:1992ws,Bagger:1997gg,Ciuchini:1997bw,Ciuchini:1998ix,Becirevic:2004qd}
that the parameters $(\delta^{d}_{12})_{IJ}$ used in our
calculations also yield important contributions to the
$K^0-\bar{K}^0$ mixing, however, our calculations will show that the
measurements of the $K_L - K_S$ mass difference and the indirect CP
violation observable $\epsilon_K$ do not lead to more stringent
constraints than ones from kaon decays.

We organize our paper as follows. In Sec.~\ref{sec:NMSSM} we give a
brief summary of the NMSSM. In Sec.~\ref{sec:expression} we
calculate the two-quark flavor-changing contributions to the
$\Sigma$ and kaon decays arising from the SUSY-FCNC effects mediated
by neutralino and gluino. In Sec.~\ref{sec:num} we combine our
two-quark contributions with four-quark contributions in
Ref.\cite{He:2006uu} to give a numerical results and discussion.
Feynman rules and analytical expressions for the four-quark
contributions are collected in the Appendix A and B, respectively.

\section{NMSSM}
\label{sec:NMSSM}

In order to make our paper self-contained, we start with a brief
description of the NMSSM and the relevant couplings considered in
our paper. The superpotential of the NMSSM is given
by\cite{Drees:1988fc,Ellis:1988er,Franke:1995tc}
\begin{equation}
W=QY_uH_uU+QY_dH_dD+LY_eH_dE+\lambda H_dH_uN-\frac{1}{3}kN^3,
\end{equation}
where $H_u$ and $H_d$ are the $SU(2)$ doublet with the hypercharge
1/2 and -1/2 and are responsible for the up- and down-type quark
mass, respectively. The ratio of the vacuum expectation values(VEVs)
of the $H_u$ and $H_d$ is defined as $\tan \beta$, which are just
like those in the the minimal supersymmetric standard model(MSSM).
Compared with the MSSM, there is one more gauge-singlet Higgs Field
$N$ with the hypercharge 0 and VEV $x$ in the NMSSM. After breaking
of the supersymmetry, there are seven physical Higgs bosons in the
NMSSM, including two charged Higgs bosons, three neutral scalar and
two pseudoscalar Higgs bosons.

The Higgs potential of NMSSM is\cite{Hiller:2004ii}
\begin{equation}
V_{\text{Higgs}}=V_{\text{soft}}+V_F+V_D,
\end{equation}
where
\begin{eqnarray}
V_{\text{soft}}&=&m^2_{H_d}|H_d|^2+m^2_{H_u}|H_u|^2-(\lambda
A_{\lambda}H_dH_uN-\frac{1}{3}kA_kN^2+\mathrm{H.c.}), \nonumber \\
V_F&=&|\lambda|^2(|H_d|^2+|H_u|^2)|N|^2+|\lambda H_dH_u-kN^2|^2,
\nonumber\\
V_D&=&\frac{g^2+g'^2}{8}(|H_d|^2-|H_u|^2)+\frac{g^2}{2}|H^{\dagger}_uH_d|^2.
\end{eqnarray}
The above Higgs potential has a global $U(1)_R$ symmetry in the
limit of vanishing parameters $A_k$, $A_{\lambda} \rightarrow
0$\cite{Dobrescu:2000yn}. If the global $U(1)_R$ symmetry is broken
slightly, the lighter pseudoscalar $A^0_1$ has a natural small mass:
\begin{equation}
m^2_{A^0_1}=3kxA_{\lambda}+\mathcal{O}\left(\frac{1}{\tan
\beta}\right),
\end{equation}
In the large $\tan \beta$ limit, $m_{A^0_1}$ can be as low as $\sim$
100MeV\cite{Hiller:2004ii}\cite{Dobrescu:2000yn}.

The Lagrangian describing the couplings of $A^0_1$ to the up- and
down-type quarks and to the leptons are given by \cite{He:2006fr}
\begin{equation}
\label{formula:Lagrangian}
\mathcal{L}_{\mathcal{A}qq}=-(l_um_u\bar{u}\gamma_5u+l_dm_d\bar{d}\gamma_5d)\frac{iA^0_1}{v}
\end{equation}
and
\begin{equation}
\mathcal{L}_{\mathcal{A}\ell}=\frac{ig_{\ell}m_{\ell}}{v}\bar{\ell}\gamma_5\ell
A^0_1,
\end{equation}
respectively, where
\begin{equation}
l_d=-g_{\ell}=\frac{v}{\sqrt{2}x}\left(\frac{A_{\lambda}-2kx}{A_{\lambda}+kx}\right),
~~~l_u=\frac{l_d}{\tan^2 \beta}.
\end{equation}
Note that $l_u$ can be neglected in the large-$\tan \beta$ limit.
The four-quark contributions can be deduced from the interactions in
Eq.~(\ref{formula:Lagrangian}) combined with the operators due to W
exchange between quarks \cite{He:2006fr}.

It has been shown in Ref.~\cite{He:2006fr} that an $A^0_1$ of mass
214.3MeV decay dominantly to muon-antimuon pair, and
$\mathcal{B}(A^0_1 \rightarrow \mu^+ \mu^-) \sim 1$ can be assumed.
In addition, the constraint imposed by the muon anomalous magnetic
moment is given by \cite{He:2005we}
\begin{equation}
|g_{\ell}|\lesssim 1.2~.
\end{equation}

Moreover, the neutralino sector of the NMSSM is different from that
in the MSSM. There are five neutralinos in the NMSSM, and the
Lagrangian for the mass term of the neutralinos can be written
as\cite{Franke:1995tc}
\begin{equation}
\mathcal{L}_{m_{\chi^0}}=-\frac{1}{2}(\psi^0)^TY\psi^0+H.c.,
\end{equation}
where Y is the symmetric neutralino mixing matrix ,its expression
can be found in Ref.~\cite{Franke:1995tc}. The masses of physical
neutralinos can be obtained by diagonalizing Y by a unitary
$5\times5$ matrix $N$
\begin{equation}
m_{\chi^0_i}\delta_{ij}=N^{\ast}_{im}Y_{mn}N_{jn}.
\end{equation}

\section{The SUSY-FCNC effects}

\label{sec:expression}
\begin{figure}
\scalebox{0.65}{\includegraphics{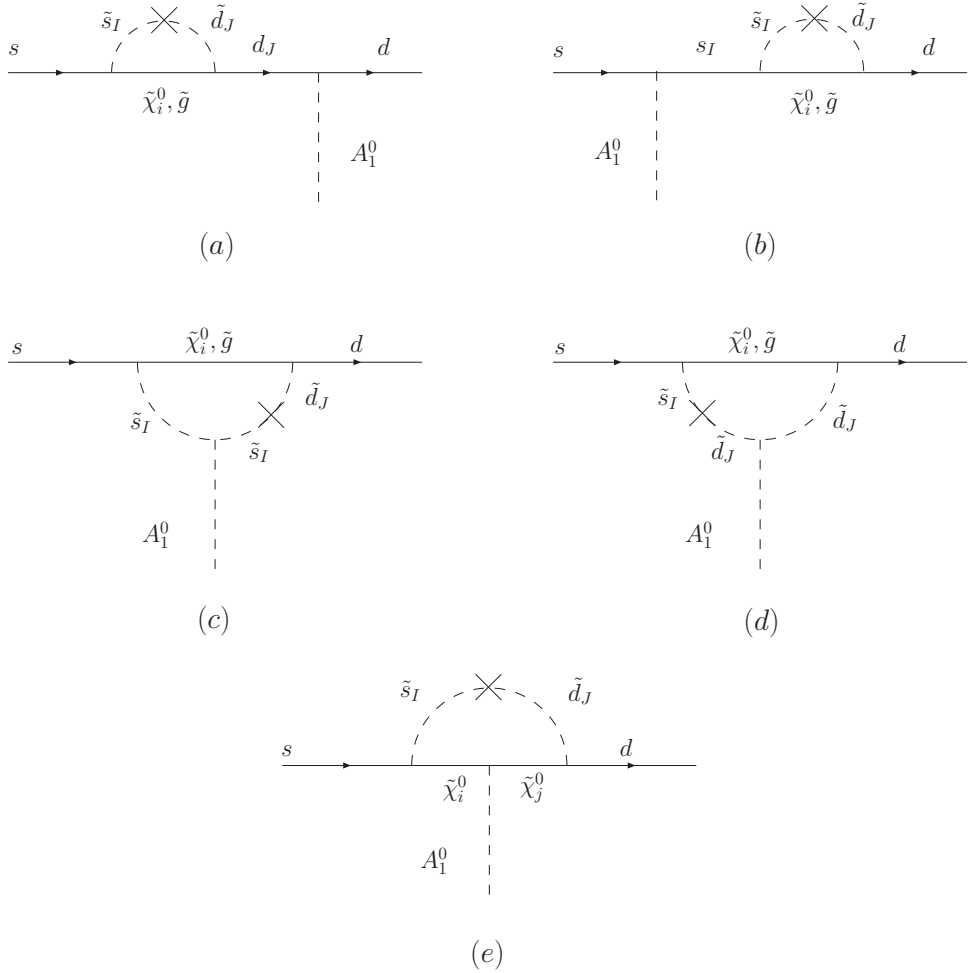}}
\caption{\label{Feynmandiagram}Feynman diagrams for $|\Delta S| =1 $
transitions, with I,J = {L,R}, and i,j = 1...5}
\end{figure}
The full hadronic amplitudes for the kaon decays in
Eq.~(\ref{formula:constraint}) and $\Sigma \to p A^0_1$ all can be
written in the following form
\begin{equation}
\mathcal{M}_{\text{full}} = \mathcal{M}_{2q} + \mathcal{M}_{4q}~,
\end{equation}
where $\mathcal{M}_{2q}$ arises from the SUSY-FCNC interactions at
the quark level, and $\mathcal{M}_{4q}$ arises from the four-quark
interactions. Using the method shown in Ref.~\cite{He:2006uu}, we
calculate the four-quark contributions, and their expressions and
numerical results are given in Appendix B. In this section, we
mainly concentrate on the hadronic amplitudes $\mathcal{M}_{2q}$
from two-quark contribution and give our analytical results. In
general, there are two kinds of FCNC contributions arising from
neutralino and gluino exchange, respectively. The relevant Feynman
diagrams for the $s \rightarrow d$ transitions are shown in Fig.
\ref{Feynmandiagram}. Note that there are five kinds of neutralinos
in the loops in the NMSSM. Calculating these Feynman diagrams, one
can obtain the two-quark FCNC Lagrangian for $s \rightarrow d A^0_1$
\begin{equation}
\mathcal{L}_{\mathcal{A}sd}=i
C_{L}\bar{d}\frac{1-\gamma_5}{2}sA^0_1+i
C_{R}\bar{d}\frac{1+\gamma_5}{2}sA^0_1+ \text{H.c.},
\label{formula:ourlagrangian}
\end{equation}
with
\begin{equation}
C_{L (R)}=C^{\tilde{\chi}^0}_{L (R)}+C^{\tilde{g}}_{L (R)},
\end{equation}
where $C^{\tilde{\chi}^0}_{L (R)}$ and $C^{\tilde{g}}_{L (R)}$
denote contributions from neutralino and gluino exchange,
respectively. They are given by
\begin{eqnarray}
C^{\tilde{\chi}^0}_L&=&\frac{\alpha}{4\pi}
\sum^5_{i,j=1}\left\{(\delta^d_{12})_{LL}\left[I^1_{ij} m_d L_{2i}
R^{\ast}_{2j} -I^3_i m_d L_{2i} R^{\ast}_{2i} + I^4_i m_d (m^2_s
L_{2i} R^{\ast}_{1i} - L_{1i} R^{\ast}_{2i}) \right]\right.\nonumber\\
&&+(\delta^d_{12})_{LR}\left[I^1_{ij} m_d m_s L_{2i} R^{\ast}_{1j}
-I^3_i m_d m_s L_{2i} R^{\ast}_{1i} - I^4_i m_d m_s (L_{1i}
R^{\ast}_{1i} + L_{2i} R^{\ast}_{2i})\right]\nonumber\\
&&+(\delta^d_{12})_{RL}\left[I^1_{ij} L_{1i} R^{\ast}_{2j} -I^3_i
L_{1i} R^{\ast}_{2i} + I^4_i (m^2_d L_{2i} R^{\ast}_{2i} + m^2_s
L_{1i} R^{\ast}_{1i})\right]\nonumber\\
&&\left.+(\delta^d_{12})_{RR}\left[I^1_{ij} m_s L_{1i} R^{\ast}_{1j}
-I^3_i m_s L_{1i} R^{\ast}_{1i} + I^4_i m_s (m^2_d
L_{2i} R^{\ast}_{1i} - L_{1i} R^{\ast}_{2i})\right]\right\},\nonumber\\
C^{\tilde{\chi}^0}_R&=&\frac{\alpha}{4\pi}
\sum^5_{i,j=1}\left\{(\delta^d_{12})_{LL}\left[I^2_{ij} m_s R_{2i}
L^{\ast}_{2j} + I^3_i m_s R_{2i} L^{\ast}_{2i} + I^4_i m_s (R_{2i}
L^{\ast}_{1i} - m^2_d R_{1i} L^{\ast}_{2i}) \right]\right.\nonumber\\
&&+(\delta^d_{12})_{LR}\left[I^2_{ij} R_{2i} L^{\ast}_{1j} + I^3_i
R_{2i} L^{\ast}_{1i} - I^4_i (m^2_d R_{1i}
L^{\ast}_{1i} + m^2_s R_{2i} L^{\ast}_{2i})\right]\nonumber\\
&&+(\delta^d_{12})_{RL}\left[I^2_{ij} m_d m_s R_{1i} L^{\ast}_{2j} +
I^3_i m_d m_s R_{1i} L^{\ast}_{2i} + I^4_i m_d m_s (R_{2i}
L^{\ast}_{2i} + R_{1i} L^{\ast}_{1i})\right]\nonumber\\
&&\left.+(\delta^d_{12})_{RR}\left[I^2_{ij} m_d R_{1i} L^{\ast}_{1j}
+ I^3_i m_d R_{1i} L^{\ast}_{1i} + I^4_i m_d (
R_{2i} L^{\ast}_{1i} - m^2_s R_{1i} L^{\ast}_{2i})\right]\right\},\nonumber\\
C^{\tilde{g}}_L&=&\frac{\alpha_s}{2\pi}C_Fm_{\tilde{g}}
[2V_{Add}C_1(y_{\tilde{g}})(\delta^d_{12})_{RL}
+V_{A\tilde{d}\tilde{d}}\frac{1}{m^2_{\tilde{d}}}D_1(y_{\tilde{g}})(m_d(\delta^d_{12})_{LL}
+m_s(\delta^d_{sd})_{RR})],
\nonumber\\
C^{\tilde{g}}_R&=&\frac{\alpha_s}{2\pi}C_Fm_{\tilde{g}}[2V_{Add}
C_1(y_{\tilde{g}})(\delta^d_{12})_{LR}-V_{A\tilde{d}\tilde{d}}\frac{1}{m^2_{\tilde{d}}}D_1(y_{\tilde{g}})
(m_d(\delta^d_{12})_{RR}+m_s(\delta^d_{sd})_{LL})],
\end{eqnarray}
with
\begin{eqnarray}
V_{Add}&=&\frac{g}{2m_W\cos\beta}U^P_{11},\nonumber\\
\label{coefficient:V}
V_{A\tilde{d}\tilde{d}}&=&\frac{g}{2m_W\cos\beta}[\lambda(v_2U^P_{13}+xU^P_{12})-A_DU^P_{11}],\\
L_{1i} &=& \frac{2 N^{\ast}_{i,1}}{3\sqrt{2} c_W},\nonumber\\
L_{2i} &=& \frac{N^{\ast}_{i,3}}{\sqrt{2} m_W s_W \cos
\beta},\nonumber\\
R_{1i} &=& L^{\ast}_{2i},\nonumber\\
\label{coefficient:LR}
R_{2i} &=& \frac{s_W N_{i,1} - 3 c_W N_{i,2}}{3\sqrt{2} c_W s_W},\\
I^1_{ij} &=&  D_2(y_i,y_j)
R^{R''}_{1ij} + \frac{m_{\tilde{\chi}_i} m_{\tilde{\chi}_j}}{m^2_{\tilde{d}}} D_2(y_i,y_j)R^{L''}_{1ij},\nonumber\\
I^2_{ij} &=&  D_2(y_i,y_j)
R^{L''}_{1ij} + \frac{m_{\tilde{\chi}_i} m_{\tilde{\chi}_j}}{m^2_{\tilde{d}}} D_2(y_i,y_j)R^{R''}_{1ij},\nonumber\\
I^3_{i} &=& 2 V_{Add} m_{\tilde{\chi}_i} C_1(y_i),\nonumber
\end{eqnarray}
\begin{eqnarray}
I^4_{i} &=&
V_{A\tilde{d}\tilde{d}}\frac{m_{\tilde{\chi}_i}}{m^2_{\tilde{d}}}
 D_1(y_i),\\
y_{\tilde{g}} &=&
\frac{m^2_{\tilde{g}}}{m^2_{\tilde{d}}},~~~y_i=\frac{m^2_{\tilde{\chi}^0_i}}{m^2_{\tilde{d}}},~i=1...5,
\end{eqnarray}
where $U^P$ is used to diagonalize the pseudoscalar Higgs mass
matrices, the matrix $N_{ij}, i,j = 1 \cdots 5$ is the $5 \times 5$
unitary matrix defined in Sec.\ref{sec:NMSSM}, $c_W = \cos
\theta_W$, where $\theta_W$ is the Weinberg angle as usual, and
\begin{eqnarray}
\label{coefficient:RLR} R^{''R}_{1ij} &=& \frac{1}{2}
\left[(U^P_{11} \cos \beta + U^P_{12}
\sin \beta) \right.\nonumber\\
&&\times\left(\frac{g}{c_W}(N_{i2}N^{\ast}_{j3}+N_{j2}N^{\ast}_{i3})
- \sqrt{2} \lambda (N_{i5}N^{\ast}_{j4} +
N_{j5}N^{\ast}_{i4})\right)\nonumber\\
&&+ (U^P_{11} \sin \beta - U^P_{12} \cos \beta)\nonumber\\
&&\left.\left(\frac{g}{c_W}(N_{i2}N^{\ast}_{j4}+N_{j2}N^{\ast}_{i4})
+ \sqrt{2} \lambda (N_{i5}N^{\ast}_{j3} +
N_{j5}N^{\ast}_{i3})\right)\right]\nonumber\\
&&-\sqrt{2}kU^P_{13}(N_{i5}N^{\ast}_{j5}+N_{j5}N^{\ast}_{i5}),\nonumber\\
R^{''R}_{1ij} &=& - R^{''L\ast}_{1ij}.
\end{eqnarray}
And the loop functions $C_1(x)$, $D_1(x)$ and $D_2(x,y)$ are defined
as
\begin{eqnarray}
C_1(x)&=&\frac{x-1-x \log(x)}{(1-x)^2},\nonumber\\
D_1(x)&=&\frac{1-x^2+2x\log(x)}{2(1-x)^3},\nonumber\\
D_2(x,y)&=&-\frac{1}{(1-x)(1-y)}-\frac{x\log(x)}{(1-x)^2(x-y)}-\frac{y\log(y)}{(1-y)^2(y-x)}.
\end{eqnarray}

The quark level effective Lagrangian in
Eq.(\ref{formula:ourlagrangian}) can be mapped onto the chiral
Lagrangian at the leading order\cite{He:2006uu}
\begin{eqnarray}
\mathcal{L}_{\mathcal{A}} &=& b_D \langle \bar{B} \{h_A, B\} \rangle
+ b_F \langle \bar{B} [h_A, B] \rangle + b_0 \langle h_A \rangle
\langle\bar{B} B\rangle + \frac{1}{2} f^2_{\pi} B_0 \langle h_A
\rangle + \text{H.c}. ,
\end{eqnarray}
where
\begin{displaymath}
B = \left( \begin{array}{ccc} \frac{1}{\sqrt{2}} \Sigma^0 + \frac{1}{\sqrt{6}} \Lambda & \Sigma^+ &p\\
 \Sigma^- & -\frac{1}{\sqrt{2}} \Sigma^0 + \frac{1}{\sqrt{6}} \Lambda &n\\
\Xi^- & \Xi^0 & -\frac{2}{\sqrt{6}}\Lambda
\end{array}\right)
\end{displaymath}
represents the baryon fields, $f_{\pi} = 92.4$MeV is the pion decay
constant, $\langle \cdots \rangle \equiv \text{Tr} ( \cdots )$ in
flavor-$SU(3)$. And
\begin{equation}
h_A = - i (C_R \xi^\dagger h \xi^\dagger + C_L \xi h\xi )A^0_1
\end{equation}
where
\begin{displaymath}
h = T_6 + i T_7 = \left( \begin{array}{ccc} 0 & 0 &0\\0 & 0 &1\\ 0 &
0 & 0
\end{array}\right)
\end{displaymath} is used to specify the $s \to d$ transition,
\begin{equation}
\xi = e^{i \pi / f_{\pi}},~~\Sigma = \xi\xi = e^{2 i \pi/f_{\pi}},
\end{equation}
and
\begin{displaymath}
\pi = \frac{1}{\sqrt{2}}\left( \begin{array}{ccc} \frac{1}{\sqrt{2}} \pi^0 + \frac{1}{\sqrt{6}} \eta & \pi^+ &K^+\\
 \pi^- & -\frac{1}{\sqrt{2}} \pi^0 + \frac{1}{\sqrt{6}} \eta &K^0\\
K^- & \bar{K}^0 & -\frac{2}{\sqrt{6}}\eta
\end{array}\right)
\end{displaymath}
is the pion octet.

The two-quark amplitude $\mathcal{M}_{2q}$ can be deduced from
$\mathcal{L}_{\mathcal{A}}$ plus the usual chiral Lagrangian
$\mathcal{L}_s$ for the strong interactions of hadrons, which is
expressed
as\cite{Gasser:1983yg,Bijnens:1985kj,Jenkins:1991bt,Jenkins:1991ne}
\begin{eqnarray}
\label{Ls} \mathcal{L}_s &=& i \langle\bar{B}\gamma^{\mu} D_{\mu} B
\rangle - m_0 \langle \bar{B} B \rangle + D \langle \bar{B}
\gamma^{\mu} \gamma_{5} \{A_{\mu}, B\} \rangle + F \langle \bar{B}
\gamma^{\mu} \gamma_{5} [A_{\mu}, B] \rangle \nonumber \\ &&+ b_D
\langle \bar{B} \{M_+, B\} \rangle + b_F \langle \bar{B} [M_+, B]
\rangle + b_0 \langle M_+ \rangle \langle \bar{B} B \rangle
\nonumber
\\ &&+\frac{1}{4}f^2_{\pi} \langle \partial^{\mu} \Sigma^{\dagger}
\partial_{\mu} \Sigma \rangle + \frac{1}{2} f^2_{\pi} B_0 \langle M_+
\rangle,
\end{eqnarray}
with
\begin{equation}
D^{\mu} B = \partial ^{\mu} B + [V^{\mu}, B], \nonumber
\end{equation}
\begin{equation}
A^{\mu} = \frac{i}{2} (\xi
\partial^{\mu} \xi^{\dagger} - \xi^{\dagger} \partial^{\mu}
\xi),~~~M_+ = \xi^{\dagger} M \xi^{\dagger} +\xi M^{\dagger}
\xi,\nonumber
\end{equation}
\begin{equation}
V^{\mu} = \frac{1}{2} (\xi \partial^{\mu} \xi^{\dagger} +
\xi^{\dagger}
\partial^{\mu} \xi), \nonumber
\end{equation}
where  $M=\text{diag}(\hat{m}, \hat{m}, m_s)$ is the quark mass
matrix in the $m_u = m_d = \hat{m}$ limit.

Using the mass relations $m_{\Sigma} - m_p =  2 (b_D - b_F)(m_s -
\hat{m})$, $m^2_K - m^2_{\pi} = B_0 (m_s - \hat{m})$ and $m^2_K =
B_0 (m_s + \hat{m})$ , the amplitudes for the different decay modes
can be written as \cite{He:2006uu}
\begin{eqnarray}
\mathcal{M}_{2q}(K^+ \to \pi^+ A^0_1) &=& -\sqrt{2}
\mathcal{M}_{2q}(K^0 \to \pi^0 A^0_1) \nonumber\\ &=& i
\left(\frac{C_L +
C_R}{2}\right) B_0,\\
\mathcal{M}_{2q}(\Sigma^+ \to p A^0_1) &=& i \left(\frac{C_L +
C_R}{2}\right)\frac{B_0 (m_{\Sigma}-m_{p})}{m^2_{K}-m^2_{\pi}}
\bar{p} \Sigma^+ \nonumber\\&&+ i (D - F)\left(\frac{C_L -
C_R}{2}\right) \frac{B_0 (m_{\Sigma} + m_{p})}{m^2_K - m^2_{A^0_1}}
\bar{p} \gamma_5 \Sigma^+,
\end{eqnarray}
where $B_0=2031$MeV\cite{He:2005we}.

\section{Numerical results and discussion}
\label{sec:num}

In this section we present our numerical results. The Higgs sector
of NMSSM is describe by the six independent parameters
\begin{equation}
\lambda,~~k,~~A_{\lambda},~~A_{k},~~\tan \beta,~~\mu, \nonumber
\end{equation}
where $\mu = -\lambda x$. For convenience, we will take $m_{A^0_1}$
and the coupling of down-type quarks to $A^0_1$, $l_d$ instead of k
and $\lambda$. We follow Ref.\cite{He:2006fr} to set $l_d=0.35$,
$-\lambda x =150$GeV and $\tan \beta = 30$. $A_k$ and $A_{\lambda}$
are set as 0.001 and 0.002, respectively. The mass of $A^0_1$ is set
as $214.3$MeV to satisfy the HyperCP data. We set the mass of gluino
and average down type squark mass as 200GeV and 350GeV,
respectively. With our input, the mass of the neutralinos are around
$100 \sim 800$GeV. Numerically, contributions from the exchange of
the gluino and squarks are lager than contributions from the
exchange of the neutralinos and squarks, i.e.,
$C^{\tilde{\chi^0}}_{L, R}$ are larger than $C^{\tilde{g}}_{LR}$ in
most regions of parameter space. This is due to the effects of
$\alpha_s$.

\begin{figure}
\includegraphics[scale=0.6]{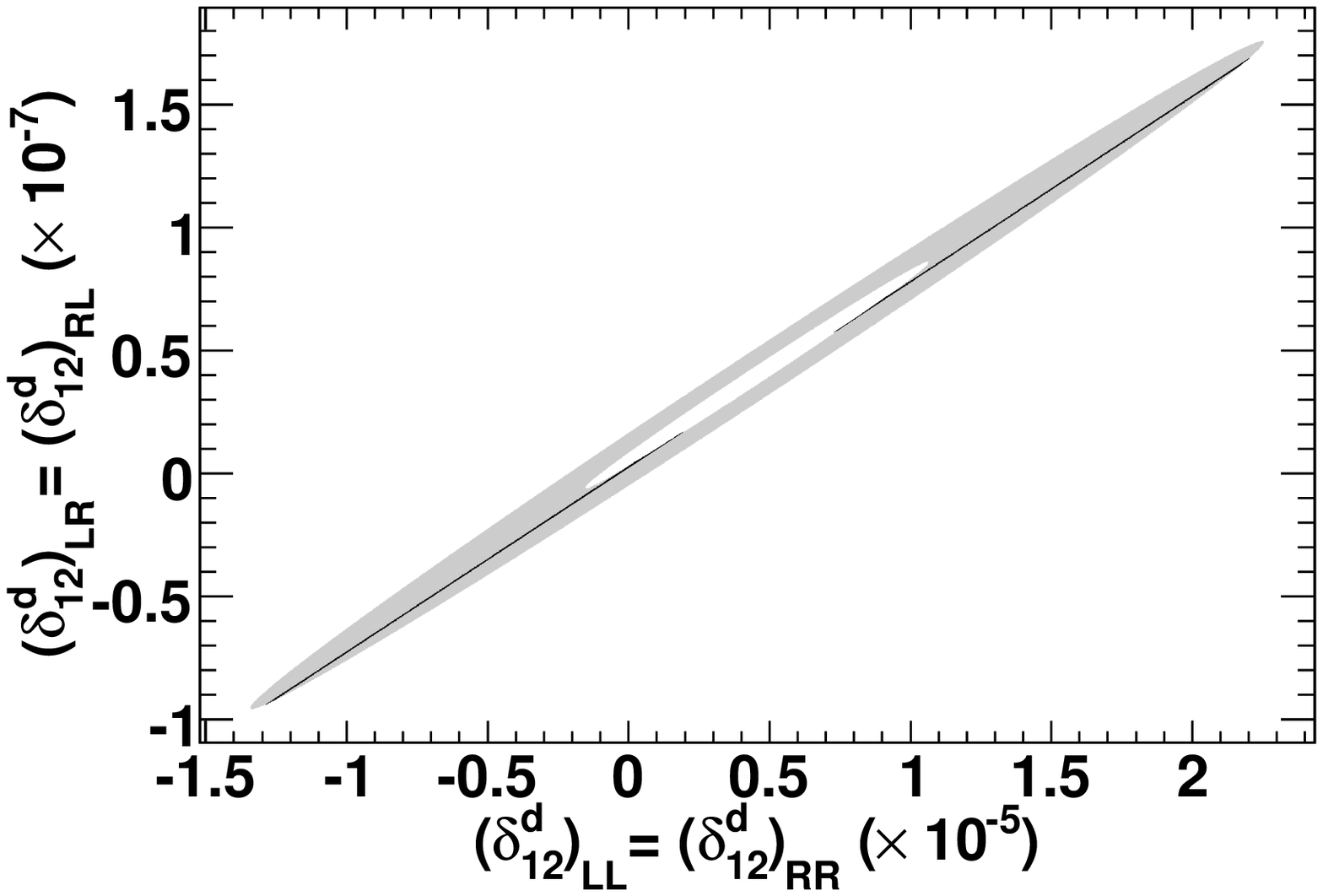}
\caption{\label{dgrm:doub}The allowed values of
$(\delta^d_{12})_{LR}=(\delta^d_{12})_{RL}$ as a function of
$(\delta^d_{12})_{LL}=(\delta^d_{12})_{RR}$. The grey area is the
regions where $A^0_1$ can explain the HyperCP events, when the
constraints from rare kaon decays are considered, the allowed
regions are greatly reduced to the dark ones.}
\end{figure}

\begin{figure}
\includegraphics[scale=0.6]{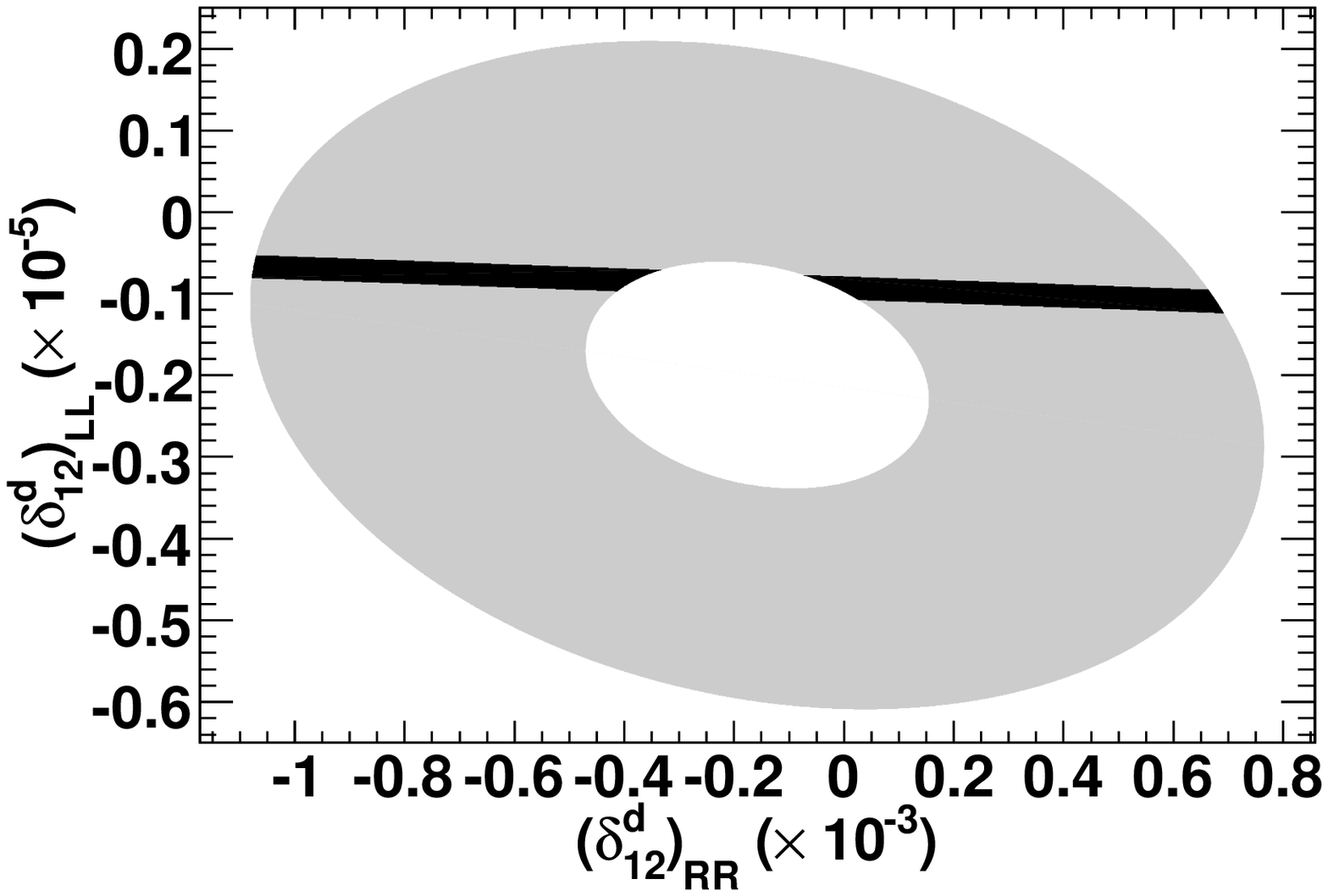}
\caption{\label{dgrm:RRLL}The allowed values of
$(\delta^d_{12})_{LL}$ as a function of $(\delta^d_{12})_{RR}$, The
grey area is the regions where $A^0_1$ can explain the HyperCP
events, when the constraints from rare kaon decays are considered,
the allowed regions are greatly reduced to the dark ones.}
\end{figure}

\begin{figure}
\includegraphics[scale=0.6]{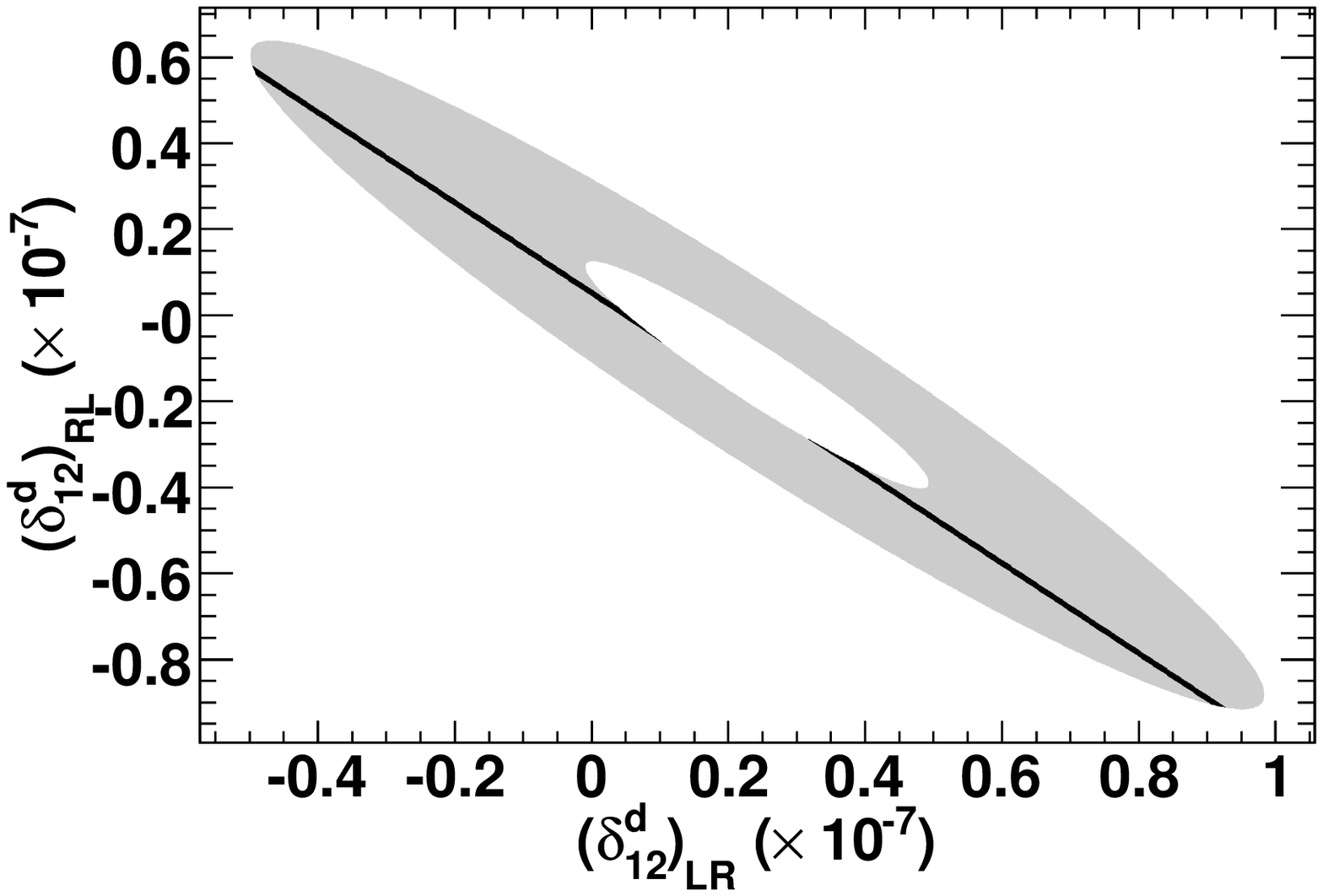}
\caption{\label{dgrm:LRRL}The allowed values of
$(\delta^d_{12})_{RL}$ as a function of $(\delta^d_{12})_{LR}$, The
grey area is the regions where $A^0_1$ can explain the HyperCP
events, when the constraints from rare kaon decays are considered,
the allowed regions are greatly reduced to the dark ones.}
\end{figure}

\begin{figure}
\includegraphics[scale=0.6]{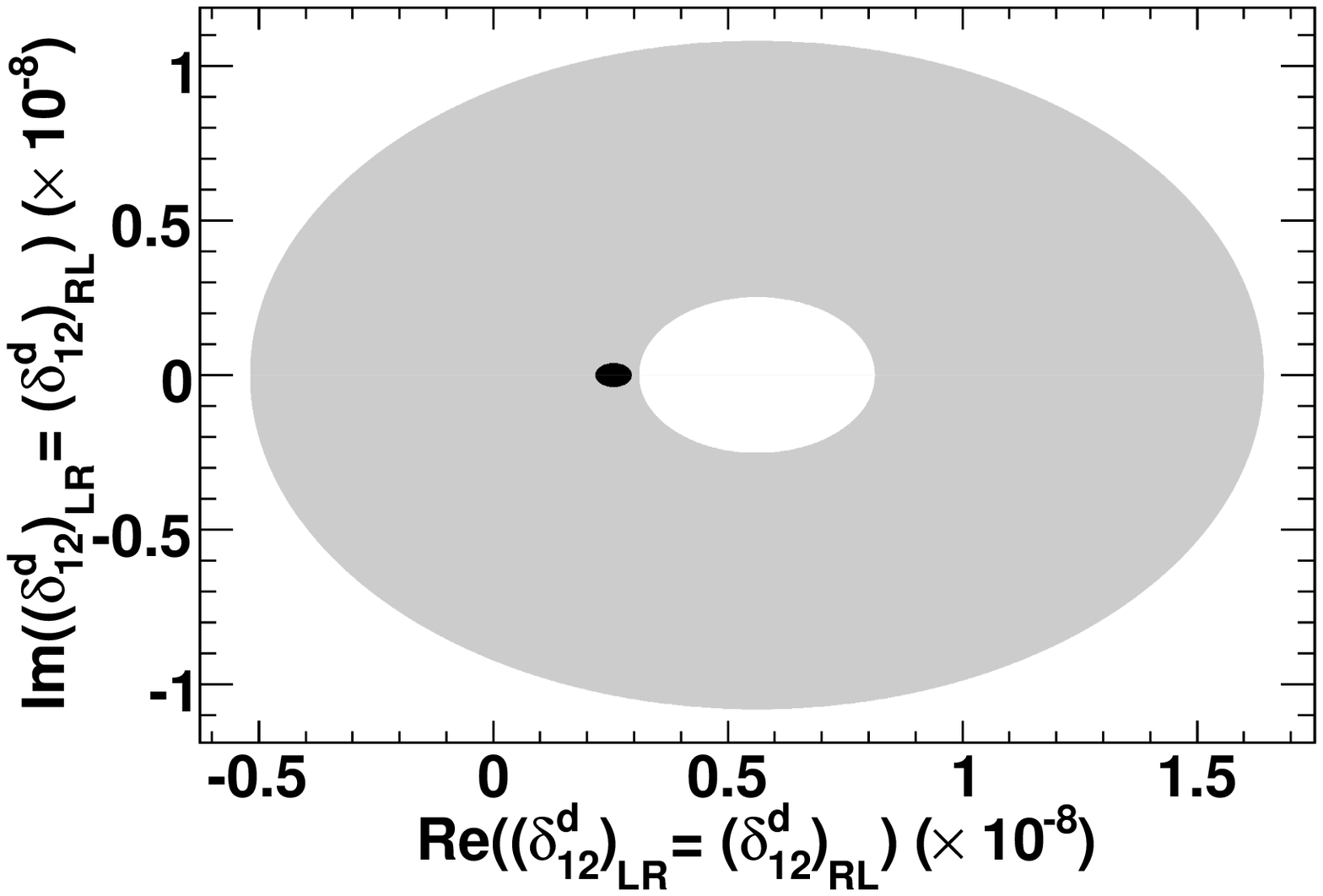}
\caption{\label{dgrm:LRRLcom}The allowed values of
$\text{Im}\{(\delta^d_{12})_{LR}=(\delta^d_{12})_{RL}\}$ as a
function of
$\text{Re}\{(\delta^d_{12})_{LR}=(\delta^d_{12})_{RL}\}$. The grey
regions denotes the survival regions for explaining the HyperCP
events alone; In the dark regions, the $A^0_1$ can explain the
HyperCP events and simultaneously satisfy the bounds originate kaon
decays and the $K^0 - \bar{K}^0$ mixing.}
\end{figure}

\begin{figure}
\includegraphics[scale=0.6]{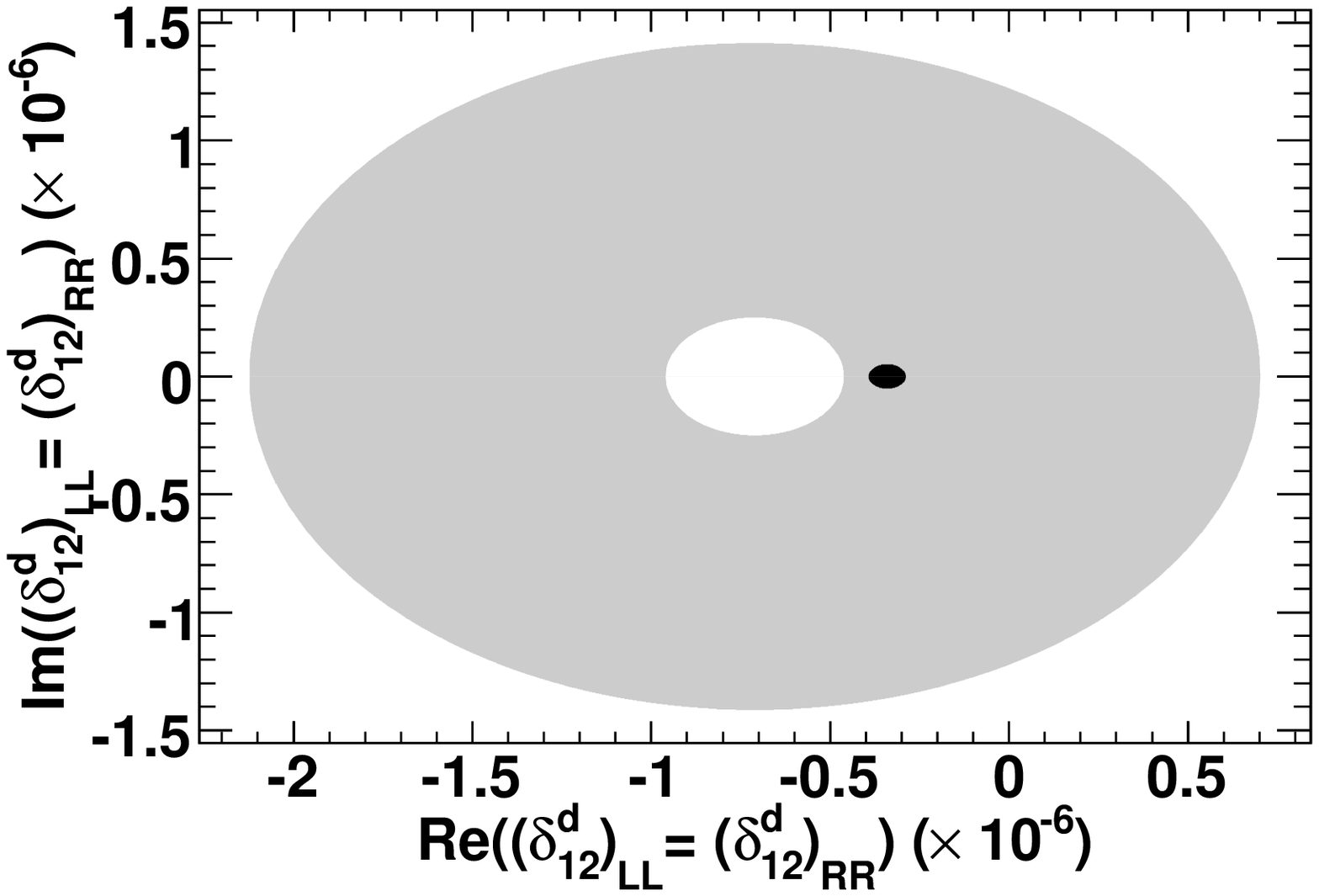}
\caption{\label{dgrm:LLRRcom}The allowed values of
$\text{Im}\{(\delta^d_{12})_{LL}=(\delta^d_{12})_{RR}\}$ as a
function of
$\text{Re}\{(\delta^d_{12})_{LL}=(\delta^d_{12})_{RR}\}$. The grey
regions denotes the survival regions for explaining the HyperCP
events alone; In the dark regions, the $A^0_1$ can explain the
HyperCP events and simultaneously satisfy the bounds originate kaon
decays and the $K^0 - \bar{K}^0$ mixing.}
\end{figure}

Our numerical results are shown in Figs.\ref{dgrm:doub} -
\ref{dgrm:LLRRcom}. We first assume that $(\delta^{12})_{IJ}$ are
real. The allowed regions in parameter space are shown in
Figs.\ref{dgrm:doub} - \ref{dgrm:LRRL}, where the grey areas are the
allowed regions for $A^0_1$ to explain the HyperCP events. When the
constraints obtained from the kaon decays are considered, the
allowed parameter space are greatly reduced to the dark regions.
From Figs.\ref{dgrm:doub} - \ref{dgrm:LRRL}, the constraints on the
combinations of the parameters can be obtained as following:
\begin{eqnarray}
\label{results:real}
(\delta^d_{12})_{LL(RR)}(\delta^d_{12})_{LR(RL)} &\le& 3.9 \times
10^{-12}~~(\text{without kaon bounds}),\nonumber\\
&& 3.7 \times 10^{-12}~~(\text{with kaon bounds}),\nonumber\\
(\delta^d_{12})_{LL}(\delta^d_{12})_{RR} &\le& 3.9 \times
10^{-9}~~~(\text{without kaon bounds}),\nonumber\\
&& 1.8 \times 10^{-14}~~(\text{with kaon bounds}),\nonumber\\
(\delta^d_{12})_{LR}(\delta^d_{12})_{RL} &\le& 2.7 \times
10^{-16}~~(\text{without kaon bounds}),\nonumber\\
&& 0.9 \times 10^{-17}~~(\text{with kaon bounds}).
\end{eqnarray}

It has been widely studied in the literature that the SUSY-FCNC
effects has great impact on the $K^0 - \bar{K}^0$ mixing if the
relevant $(\delta^d_{12})_{IJ}$ are complex. So we further
investigate the possible constraints on $(\delta^d_{12})_{IJ}$ from
the $K_L - K_S$ mass difference $\Delta m_K$ and indirect CP
violation parameter $\epsilon_K$. However, the constraints shown in
Eq.(\ref{results:real}) are roughly several orders smaller than
those given in the literatures involving the SUSY-FCNC mediated $K^0
- \bar{K}^0$ mixing
\cite{Hagelin:1992tc,Gabbiani:1996hi,Hagelin:1992ws,Bagger:1997gg,Ciuchini:1997bw,Ciuchini:1998ix,Becirevic:2004qd},
where $(\delta^d_{12})_{LL}$ and $(\delta^d_{12})_{LR(RL)}$ are
around $\mathcal{O}(10^{-1} \sim 10^{-3})$ and $\mathcal{O}(10^{-3}
\sim 10^{-4})$, respectively. This fact indicates that the
constraints from the $K^0 - \bar{K}^0$ mixing may be automatically
satisfied once the constraints from $\Sigma \to p A^0_1$ and the
rare kaon decays in Eq.(\ref{formula:constraint}) are satisfied. Our
numerical results do indeed confirm that the constraints from the
$K^0 -\bar{K}^0$ mixing do not lead to more stringent constraints
than those ones given from the kaon decays in
Eq.(\ref{formula:constraint}) .

Figs.\ref{dgrm:LRRLcom} and \ref{dgrm:LLRRcom} show the constraints
on the complex $(\delta^d_{12})_{IJ}$ from $\Sigma \to p A^0_1$ and
rare kaon decays in Eq.(\ref{formula:constraint}). And the
corresponding constraints on the combinations of parameters are
given by
\begin{eqnarray}
\text{Re}(\delta^d_{12})_{LR(RL)}\text{Im}(\delta^d_{12})_{LR(RL)}
&\le& 2.7 \times
10^{-16}~~(\text{without kaon bounds}) \nonumber,\\
&& 1.5 \times 10^{-18}~~(\text{with kaon bounds}),\nonumber\\
\text{Re}(\delta^d_{12})_{LL(RR)}\text{Im}(\delta^d_{12})_{LL(RR)}
&\le& 1.8 \times
10^{-12}~~(\text{without kaon bounds}),\nonumber\\
&& 1.8 \times 10^{-14}~~(\text{with kaon bounds}).
\end{eqnarray}
From Figs.\ref{dgrm:LRRLcom} and \ref{dgrm:LLRRcom}, it can be seen
that the grey areas are the allowed regions of the SUSY-FCNC
parameters for $A^0_1$ to explain the HyperCP events and the grey
regions are greatly reduced to the dark ones when the constraints
from rare kaon decays are considered. Even so, there are still
regions in the SUSY-FCNC parameter space where $A^0_1$ in the NMSSM
can be used to explain the HyperCP events without contradicting with
the constraints from the rare kaon decays and the $K^0 - \bar{K}^0$
mixing.

In conclusion, we have calculated the two-quark contributions to the decay $\Sigma^+ \to p \mu^+ \mu^-$ arising from
the transition $s \to d A^0_1$ via the SUSY-FCNC couplings. Combining the two-quark contributions with the
four-quark contributions, we show that there
are regions in the SUSY-FCNC parameter space where the $A^1_0$ in the NMSSM
can be identified with a new particle of mass 214.3MeV, X, which can be used to explain the HyperCP events,
while satisfying all the constraints from the measurements of the rare
kaon decays. And once the constraints from the kaon decays are
satisfied, the constraints from the $K^0-\bar{K}^0$ mixing are
automatically satisfied.

\begin{acknowledgments}
This work was supported in part by the National Natural Science
Foundation of China, under Grant No.~10421503, No.~10575001 and
No.~10635030, and the Key Grant Project of Chinese Ministry of
Education under Grant No.~305001.
\end{acknowledgments}

\section*{Appendix A}

\begin{figure}
\includegraphics[scale=0.7]{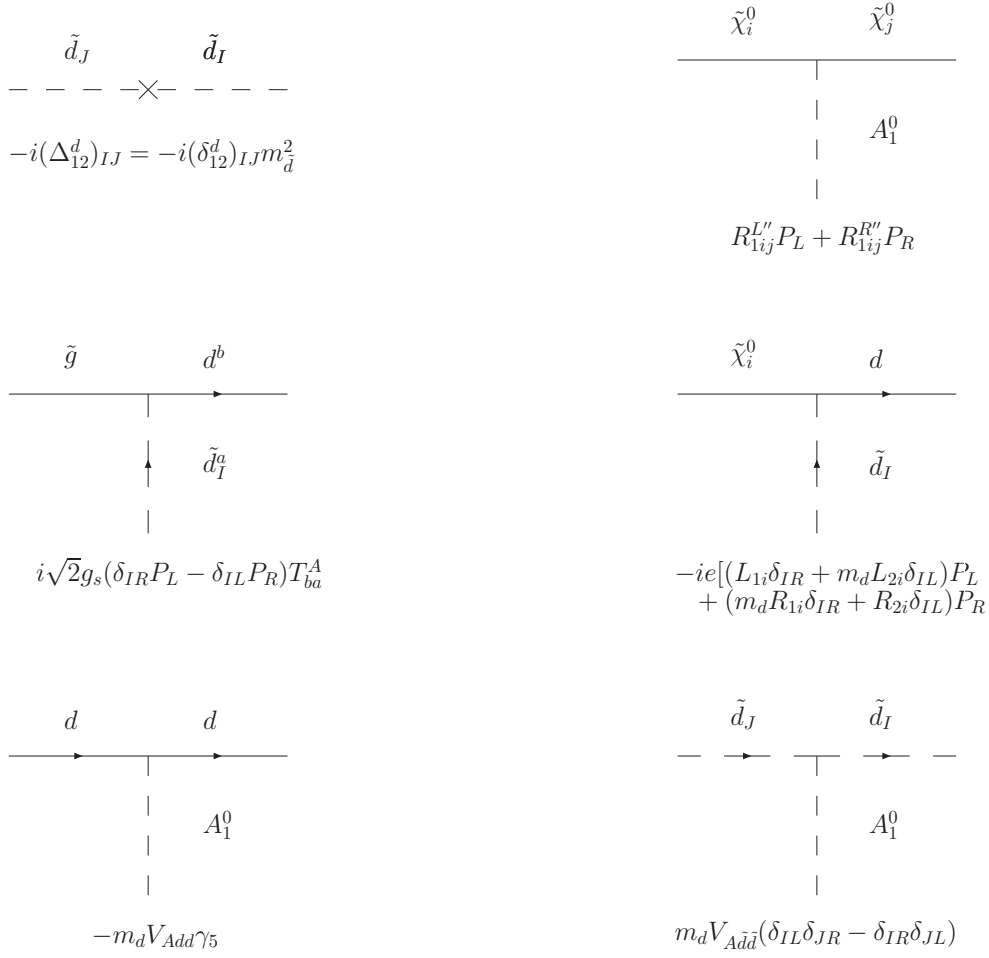}
\caption{\label{feynmanrule}Feynman rules used in our paper.}
\end{figure}

We give the Feynman rules used in our calculations in
Fig.\ref{feynmanrule}, where $V_{A\tilde{d}\tilde{d}}$, $V_{Add}$,
$L(R)_{1(2)i}$ and $R^{L(R)''}_{1ij}$ are defined in Eq.
(\ref{coefficient:V}), Eq. (\ref{coefficient:LR}) and Eq.
(\ref{coefficient:RLR}), respectively. And
\begin{displaymath}
\delta_{IJ} = \left\{ \begin{array}{ll} 1 & ~~I=J, \\
0 & ~~I \neq J .
\end{array} \right.
\end{displaymath}

\section*{Appendix B}

We collect expressions of the four-quark amplitudes for the
different decay modes in this appendix. A detailed description can
be found in Ref. \cite{He:2006uu}, we cite their results here. As
pointed in Ref. \cite{He:2006fr}, the couplings of $A^0_1$ to the
up-type quarks tends to zero at the limit of large tan$\beta$, so we
neglect terms that are proportional to $l_u$.

The four-quark contributions for the kaon decays are as follow:
\begin{eqnarray}
\mathcal{M}_{4q}(K^+ \to \pi^+ A^0_1) & = & i \frac{l_d \gamma_8}{v}
\left\{ -\frac{ m^2_{\pi}}{2} + [(2 m^2_K + m^2_{\pi} - 3
m^2_{A^0_1}) c_{\theta} - \sqrt{8} (m^2_K - m^2_{\pi})
s_{\theta}]\right.\nonumber\\ && \times \frac{(4 m^2_K - 3
m^2_{\pi}) c_{\theta} + \sqrt{2} (2 m^2_K -\tilde{m}^2_0)
s_{\theta}}{6 (m^2_{\eta} - m^2_{A^0_1})} + \left[(2 m^2_K +
m^2_{\pi} - 3 m^2_{A^0_1}) s_{\theta} \right.\nonumber\\
&&\left.\left. + \sqrt{8} (m^2_K - m^2_{\pi}) c_{\theta}\right]
\frac{(4 m^2_K - 3 m^2_{\pi}) s_{\theta} - \sqrt{2} (2 m^2_K
-\tilde{m}^2_0) c_{\theta}}{6 (m^2_{\eta^{\prime}} -
m^2_{A^0_1})}\right\},\\
\mathcal{M}_{4q}(K^0 \to \pi A^0_1) & = & i \frac{l_d \gamma_8}{v}
\left\{ -\frac{(2 m^2_K - m^2_{\pi} -m^2_{A^0_1})
m^2_{\pi}}{\sqrt{8} (m^2_{A^0_1} - m^2_{\pi})} \right.\nonumber\\&&+
[(2 m^2_K + m^2_{\pi} - 3 m^2_{A^0_1}) c_{\theta} - \sqrt{8} (m^2_K
- m^2_{\pi}) s_{\theta}]\nonumber\\ && \times \frac{(4 m^2_K - 3
m^2_{\pi}) c_{\theta} + \sqrt{2} (2 m^2_K -\tilde{m}^2_0)
s_{\theta}}{6 \sqrt{2} ( m^2_{A^0_1} - m^2_{\eta})} + \left[(2 m^2_K
+ m^2_{\pi} - 3 m^2_{A^0_1}) s_{\theta} \right.\nonumber\\
&&\left.\left. + \sqrt{8} (m^2_K - m^2_{\pi}) c_{\theta}\right]
\frac{(4 m^2_K - 3 m^2_{\pi}) s_{\theta} - \sqrt{2} (2 m^2_K
-\tilde{m}^2_0) c_{\theta}}{6 \sqrt{2} (m^2_{A^0_1} -
m^2_{\eta^{\prime}})}\right\},
\end{eqnarray}
where $\gamma_8 = -7.8 \times 10^{-8}$, $s_{\theta}$ and
$c_{\theta}$ are short for $\sin \theta$ and $\cos \theta$, $\theta
= -19.7^{\circ}$. And, the four-quark contributions to the $\Sigma
\to p A^0_1$ process are
\begin{eqnarray}
\mathcal{M}_{4q}(\Sigma^+ \to p A^0_1)  = i \bar{p} (A_{p A^0_1} -
B_{p A^0_1} \gamma_5) \Sigma^+,
\end{eqnarray}
with
\begin{eqnarray}
A_{p A^0_1} & = & l_d \frac{f_{\pi}}{v} \frac{A_{p \pi^0}}{2}\left\{
\frac{m^2_{\pi}}{m^2_{A^0_1} - m^2_{\pi}} + \frac{(4 m^2_K - 3
m^2_{\pi}) c^2_{\theta} + \sqrt{2} (2 m^2_K - \tilde{m}^2_0)
c_{\theta} s_{\theta}}{m^2_{\eta} - m^2_{A^0_1}}
\right.\nonumber\\&& \left.+ \frac{(4 m^2_K - 3 m^2_{\pi})
s^2_{\theta} - \sqrt{2} (2 m^2_K - \tilde{m}^2_0) c_{\theta}
s_{\theta}}{m^2_{\eta^{\prime}} - m^2_{A^0_1}} \right\}
\end{eqnarray}
and
\begin{eqnarray}
B_{p A^0_1} & = & l_d \frac{f_{\pi}}{v} \frac{B_{p
\pi^0}}{2}\left\{\frac{m^2_{\pi}}{m^2_{A^0_1} - m^2_{\pi}} +
\frac{(4 m^2_K - 3 m^2_{\pi}) c^2_{\theta} + \sqrt{2} (2 m^2_K -
\tilde{m}^2_0) c_{\theta} s_{\theta}}{m^2_{\eta} - m^2_{A^0_1}}
\right.\nonumber\\&&\left. + \frac{(4 m^2_K - m^2_{\pi})
s^2_{\theta} - \sqrt{2} (2 m^2_K - \tilde{m}^2_0) c_{\theta}
s_{\theta}}{m^2_{\eta^{\prime}} - m^2_{A^0_1}} \right\},
\end{eqnarray}
where $A_{p \pi^0} = -3.25 \times 10^{-7}$, $B_{p \pi^0} = 26.67
\times 10^{-7}$.

Numerically, the above amplitudes are
\begin{eqnarray}
\mathcal{M}_{4q}(\Sigma \to pA^0_1) &=& i\bar{p}(-6.96\times10^{-7}l_d\frac{f_{\pi}}{v}-(5.71\times10^{-6})l_d\frac{f_{\pi}}{v}\gamma_5)\Sigma^+,\nonumber\\
\mathcal{M}_{4q}(K^+ \to \pi^+ A^0_1) &=& -i1.08\times10^{-7}l_d\frac{m^2_K}{v},\nonumber\\
\mathcal{M}_{4q}(K^0 \to \pi A^0_1) &=&
i1.12\times10^{-7}l_d\frac{m^2_K}{v}.
\end{eqnarray}

\bibliography{sigma}

\end{document}